\documentclass[aps,prd,showpacs,nofootinbib,tightenlines]{revtex4}
\usepackage{amsmath}
\usepackage{amssymb}
\usepackage{epsfig}
\usepackage{graphicx}
\usepackage{hyperref}
\usepackage{color}
\usepackage{ulem}

\allowdisplaybreaks

\begin{document}

\title{Quasi-two-body decays $B\to P f_0(500)\to P\pi^+\pi^-$ in the perturbative QCD approach}

\author{Jia-Wei Zhang$^1$}
\author{Bo-Yan Cui$^1$}
\author{Xing-Gang Wu$^2$}
\author{Hai-Bing Fu$^3$}
\email{fuhb@gzmu.edu.cn}
\author{Ya-Hui Chen$^4$}
\email{cyh@tmmu.edu.cn}

\affiliation{$^1$Department of Physics, Chongqing University of Science and Technology, Chongqing 401331, People's Republic of China}
\affiliation{$^2$Department of Physics, Chongqing Key Laboratory for Strongly Coupled Physics,Chongqing University, Chongqing 401331, People's Republic of China}
\affiliation{$^3$Department of Physics, Guizhou Minzu University, Guiyang 550025, People's Republic of China}
\affiliation{$^4$Department of Physics, College of Basic Medical Sciences, Army Medical University, Chongqing 400038, People's Republic of China}

\date{\today}

\begin{abstract}
In this paper, we study the quasi-two-body decays $B\to P f_0(500)\to P\pi^+\pi^-$ [with $P=(\pi, K, \eta, \eta^{\prime})$] within framework of perturbative QCD (PQCD) factorization approach. With the help of $\pi$-$\pi$ distribution amplitude and scalar form factor $F_{\pi\pi}(\omega^2)$, we calculate the $CP$ averaged branching fraction and the $CP$ asymmetry for the quasi-two-body decays $B\to P f_0(500)\to P\pi^+\pi^-$. Taking the quasi-two-body decay $B^+ \to \pi^+ f_0(500) \to \pi^+ \pi^+ \pi^-$ as an explicit example, we present the behavior of differential branching fraction and direct $CP$ violation versus the $\pi$-$\pi$ invariant mass. The total branching fraction and direct $CP$ violation are $\mathcal{B}(B^+\to \pi^+ [\sigma\to]\pi^+\pi^-) = (1.78 \pm 0.41\pm 0.51) \times 10^{-6}$ and $\mathcal{A}_{CP}(B^+\to \pi^+ [\sigma\to]\pi^+\pi^-) = (29.8\pm 11.1\pm 13.0)\%$ respectively. Our results could be tested by further experiments.
\end{abstract}


\maketitle

\section{INTRODUCTION}

Three-body $B$-meson decays are considerably more challenging than that of two-body-decays, mainly due to the entangled resonant and nonresonant contributions, the complex interplay between the weak and strong dynamics~\cite{Charles:2017ptc}, and other possible final-state-interactions (FSI)~\cite{Bediaga:2015mia, Bediaga:2017axw, Yu:2024kjw} in the three-body $B$ meson decays. Traditional approaches for the two-body-decays are no longer satisfactory in the three-body decay processes~\cite{AlvarengaNogueira:2016cbm}. Practically, the hadronic three-body $B$ meson decay processes, in most cases,  are considered to be dominated by the low-energy $S$-, $P$- and $D$-wave resonant states, which could be treated in the quasi-two-body framework. By neglecting the FSI between the meson pair originated from the resonant states and the bachelor particle, the factorization procedure can be applied~\cite{AlvarengaNogueira:2016cbm, Boito:2017jav}. Substantial theoretical efforts for different quasi-two-body $B$ meson decays has been made within different theoretical approaches, cf. Refs.~\cite{Gronau:2003ep, Engelhard:2005hu, Gronau:2005ax, Imbeault:2011jz, Gronau:2013mda, Bhattacharya:2013cvn, Bhattacharya:2014eca, Xu:2013rua, Xu:2013dta, He:2014xha, Furman:2005xp, El-Bennich:2006rcn, El-Bennich:2009gqk, Leitner:2010ai, Dedonder:2010fg, Cheng:2005ug, Cheng:2007si, Cheng:2013dua, Cheng:2014uga, Li:2014oca, Cheng:2016shb, Krankl:2015fha, Klein:2017xti}. As well, the contributions from various intermediate resonant states for the three-body $B$-meson decays in the context of perturbative QCD (PQCD) approach~\cite{Keum:2000ph, Keum:2000wi, Lu:2000em} have been investigated in Refs.~\cite{Wang:2016rlo, Ma:2017idu, Ma:2017kec, Rui:2017hks, Li:2018lbd, Wang:2018dfq, Ma:2019qlm, Cui:2019khu, Wang:2020saq, Fan:2020gvr, Chai:2021pyp}.

Compared with vector and tensor mesons, the identification of the scalar mesons is long-standing puzzle~\cite{ParticleDataGroup:2022pth}. Scalar resonances are hard to be resolved, since some of them have large decay widths which make us difficult to distinguish between resonance and background. In the theoretical point of view, their masses do not fit the expectation in the naive quark model. For the lightest scalar meson $f_{0}(500)$ (also refereed as $\sigma$) meson, which is, in the present state, not a ordinary meson in the sense that it cannot be interpreted as predominantly made of quark and antiquark~\cite{Pelaez:2015qba}. After more than 60 years of study of $f_{0}(500)$, various of interpretations have been proposed. More explicitly, a light scalar-isoscalar field was first postulated~\cite{Johnson:1955zz} for explanation of the inter-nucleon attraction. Then, Linear Sigma Model was proposed~\cite{Gell-Mann:1960mvl} to describe the chiral symmetry in pion-pion interaction, this explains why $f_{0}(500)$ is usually called as the $\sigma$ meson. The linear sigma model plays a relevant role, in history, for the understanding of spontaneous chiral symmetry breaking, where all fields become Goldstone bosons, i.e., pions, except $\sigma$. When the $f_{0}(500)$ and $f_{0}(980)$ are considered as $q\bar{q}$ state, the mixing of light and strange quark may appear~\cite{Stone:2013eaa}, it can be characterized by a $2\times2$ rotation matrix with a single parameter, i.e., the mixing angle $\phi$. If the $q\bar{q}$ state does exist, the mixing angle can be constrained by the scalar decay channel which is exactly what we study. Comparison between our theoretical prediction and the experimental data may allow us to probe the inner structure of light scalar mesons. Tetraquarks is also a popular interpretation~\cite{Jaffe:1976ig} which is bounded to form a color neutral resonance by two valence quark and two antiquark~\cite{Pelaez:2015qba}. There are also some different explanation of quark level dynamics to describe the formation of scalar mesons, such as the bag model with additional one gluon exchange~\cite{Jaffe:1976ig}, the diquark-antidiquark configurations~\cite{Maiani:2004uc}, the large-$N$ quantum chromodynamics~\cite{Weinberg:2013cfa}, moreover, including instanton effects~\cite{tHooft:2008rus}. In this theories, tetraquark states are unmixed, as assumed in Ref.~\cite{Stone:2013eaa}, with a constrain of mixing angle less than 5 degree~\cite{Fleischer:2011au}. Consequently, how to distinguish $q\bar{q}$ and $q\bar{q}q\bar{q}$ states becomes a significant issue. In Ref.~\cite{Wang:2009azc}, authors proposed a method to distinguish two kinds of scalar mesons based on sum rule technique, it may becomes a important criterion of phenomenological study and may provide more detail of the inner structure of scalar mesons.

\begin{figure}[hbp]
\centerline{\epsfxsize=13cm \epsffile{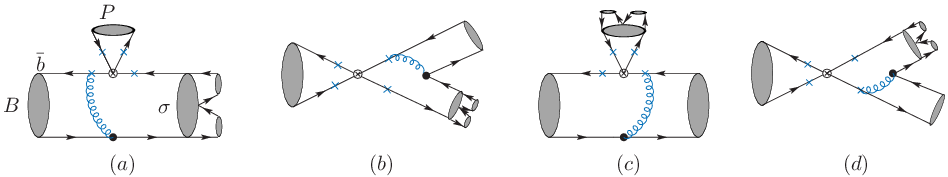}}
\vspace{0.3cm}
\caption{Typical diagrams for the quasi-two-body decays $B_{(s)}\to P\sigma\to P\pi\pi$. The diagram (a) for the $B\to \sigma$ transition, and diagram (c) for the $B\to P$ transition, as well as the diagrams (b) and (d) for annihilation contributions. The symbol $\otimes$ stands for the weak vertex and $\times$ denotes possible attachments of hard gluons.}
\label{fig-fig1}
\end{figure}

Within the framework of PQCD approach, quasi-two body decays involving resonant state $f_0(500)$ have been studied~\cite{Wang:2015uea, Ma:2017idu, Liu:2019ymi, Xing:2019xti}, we would like to extend to our previous studies to the quasi-two-body $B$ meson decays $B \to P f_0(500) \to P \pi \pi$, where the bachelor particle $P$ denotes the light pseudoscalar $\pi$, $K$, $\eta$, or $\eta^{\prime}$. Typical diagrams for the $B \to P f_0(500) \to P \pi \pi$ decay processes are shown in Fig.~\ref{fig-fig1}. Inspired by the generalized parton distribution (GPD) in hard exclusive two pion production~\cite{Diehl:1998dk, Muller:1994ses, Polyakov:1998ze, Hagler:2002nh}, the two-meson distribution amplitude was introduced in three-body hadronic $B$ decays in the frame work of PQCD approach~\cite{Chen:2002th, Chen:2004az} as the universal nonperturbative input. The decay amplitude for the quasi-two-body decays $B \to P f_0(500) \to P \pi \pi$ can be expressed as the convolution of the nonperturbative wave function and hard kernel~\cite{Chen:2002th, Chen:2004az, Wang:2016rlo}
\begin{eqnarray}
\mathcal{A}=\phi_{B} \otimes H \otimes \phi_P \otimes \phi_{\pi\pi}^{S\text{-wave}},
\end{eqnarray}
where hard kernel $H$ contains one hard gluon, and the distribution amplitudes $\phi_{B}, \phi_P$ and $\phi_{\pi\pi}^{S\text{-wave}}$ absorb the nonperturbative dynamics in the decay processes.

The rest of this paper are organized as follows. We give a brief introduction of the theoretical framework in Sec.~\ref{section2}. Numerical results and some discussions are shown in Sec.~\ref{section3}, and a brief conclusion will be summarized in Sec.~\ref{section4}. The relevant factorization formulas for the decay amplitudes are collected in the Appendix.

\section{FRAMEWORK}\label{section2}

In the light-cone coordinate, the $B$ meson momentum $p_B$, $\sigma$ meson momentum $p$, bachelor particle momentum $p_3$, and the corresponding quark momenta $k_B$, $k$, $k_3$ in the rest frame of $B$ meson are defined as
\begin{align}
&p_B=\frac{m_B}{\sqrt2}(1,1,\textbf{0}_{\rm T}),&&
k_B=\left(0,\frac{m_B}{\sqrt2}x_B ,\textbf{k}_{B{\rm T}}\right),\nonumber\\
&p=\frac{m_B}{\sqrt2}(1,\eta,\textbf{0}_{\rm T}),&&
k=\left(\frac{m_B}{\sqrt2}z,0,\textbf{k}_{\rm T}\right),\nonumber\\
&p_3=\frac{m_B}{\sqrt2}(0,1-\eta,\textbf{0}_{\rm T}),&&
k_3=\left(0,\frac{m_B}{\sqrt2}(1-\eta)x_3,\textbf{k}_{3{\rm T}}\right),
\end{align}
where the variable $\eta$ is defined as $\eta=\omega^2/m_B^2$ with $\omega=\sqrt{p^2}$ stand for invariant mass of dipion. Terms $x_B$, $z$, and $x_3$ are the momentum fraction of $k_B$, $k$, and $k_3$ respectively.

The $B$ meson can be treated as heavy-light system, whose wave function can be written as \cite{Li:2003yj}
\begin{eqnarray}
\Phi_{B}(x,b)=\frac{1}{\sqrt{2N_c}}\left({p_B}\hspace{-4.1truemm}/ \ +m_B\right)\gamma_5\phi_{B}(x,b),
\end{eqnarray}
where $b$ is the conjugate space coordinate of the transverse momentum of the valence quark of meson, $N_c$ is the color factor, the distribution amplitude $\phi_B(x,b)$ are chosen as
\begin{eqnarray}
\phi_{B}(x,b)= N_B x^2 (1-x^2)\exp\left[-\frac{1}{2}\left(\frac{x m_B}{\omega_B}\right)^2-\frac{\omega_B^2 b^2}{2}\right],
\end{eqnarray}
where the $\omega_{B}$ is the shape parameter and $N_B$ being the normalization factor. The shape parameter $\omega_B$ is mainly fixed from the fit to the $B\to\pi$ form factors derived from lattice QCD~\cite{Bowler:1999xn} and LCSR~\cite{Ball:1998tj}. The normalization constant $N_B$ is related to the decay constant $f_B$ through the relation
\begin{eqnarray}
\int^1_0 dx \phi_B(x,b=0)=\frac{f_B}{2\sqrt{6}}.
\end{eqnarray}
For the final state of light pseudoscalar meson, whose wave function being the
\begin{eqnarray}
\Phi_{P}(p,z)=\frac{i}{\sqrt{2N_c}}\gamma_5\left[p\hspace{-1.6truemm}/ \phi^A(z)+m_0\phi^P(z)+
m_0(v\hspace{-1.5truemm}/ n\hspace{-1.8truemm}/-1)\phi^T(z)\right],
\end{eqnarray}
where $m_0$ is chiral mass of corresponding pseudoscalar meson. The $p$ and $z$ are associated momentum and momentum fraction respectively. The explicit expression of relevant distribution amplitude in light-cone sum rule up to twist-3 are shown \cite{Ball:1998tj,Ball:1998je,Ball:2004ye,Ball:2006wn}
\begin{eqnarray}
\phi^A(x)&=&\frac{f_{P}}{2\sqrt{2N_c}}6x(1-x)\left[1+a_1^{P}C_1^{3/2}(2x-1)+a_2^{P}C_2^{3/2}(2x-1)+a_4^{P}C_4^{3/2}(2x-1)\right], \nonumber\\
\phi^P(x)&=&\frac{f_{P}}{2\sqrt{2N_c}}\bigg[1+\left(30\eta_3-\frac{5}{2}\rho^2_{P}\right)C_2^{1/2}(2x-1)-3\left[\eta_3\omega_3+\frac{9}{20}\rho^2_{P}(1+6a_2^{P})\right] C_4^{1/2}(2x-1)\bigg], \nonumber\\
\phi^T(x)&=&\frac{f_{P}}{2\sqrt{2N_c}}(1-2x)\bigg[1+6\left(5\eta_3-\frac{1}{2}\eta_3\omega_3-\frac{7}{20}\rho^2_{P}-
\frac{3}{5}\rho^2_{P}a_2^{P}\right)(1-10x+10x^2)\bigg],
\end{eqnarray}
with the Gegenbauer moments are $a_1^{\pi,\eta_{q,s}}=0,\ a_1^K=0.06,\ a_2^{\pi,K}=0.25,\ a_2^{\eta_{q,s}}=0.115,\ a_4^{\pi,\eta_{q,s}}=-0.015$. Meanwhile, the parameters are $\rho_{\pi}=m_{\pi}/m_0^{\pi},\ \rho_{K}=m_{K}/m_0^{K},\ \rho_{\eta_q}=2m_q/m_0^q,\ \rho_{\eta_s}=2m_s/m_0^s$, $\eta_3=0.015,\ \omega_3=-3$ with
$m_0^{\pi}=(1.4 \pm 0.1)$~GeV, $m_0^{K}=(1.6 \pm 0.1)$~GeV, $m_0^{\eta_q}=1.07$~GeV, $m_0^{\eta_s}=1.92$~GeV. The definition of Gegenbauer polynomials can be found in Ref.~\cite{Ball:2004ye,Ball:2006wn}.

For the $\eta$ and $\eta^{\prime}$ mesons, its components has been studied extensively, two major mixing mechanism were adopt in most study, the quark flavor basis and singlet-octet basis. Different process under different assumption are studied to determine the mixing angle, we prefer to choose the so-called Feldmann-Kroll-Stech (FKS) formalism~\cite{Feldmann:1998vh,Feldmann:1998sh} in which it was considered as mixing of $\eta_q$ and $\eta_s$, which made of $n\bar{n}=(u\bar{u}+d\bar{d})/\sqrt{2}$ and $s\bar{s}$ respectively, the physical state $\eta$ and $\eta^{\prime}$ related to flavor state $\eta_q$ and $\eta_s$ through a two by two rotation matrix with a single parameter, the mixing angle $\phi$
\begin{eqnarray}
\left(\begin{array}{c}\vert \eta \rangle \\ \vert \eta^{\prime} \rangle \end{array}\right)=
\left(\begin{array}{cc} \text{cos}\phi & -\text{sin}\phi \\ \text{sin}\phi & \text{cos}\phi \end{array} \right)
\left(\begin{array}{c}\vert \eta_q \rangle \\ \vert \eta_s  \rangle \end{array}\right)
\end{eqnarray}
with $\phi=39.3^{\circ} \pm 1.0^{\circ}$. For the possible glueball effect are considered to be small \cite{Charng:2006zj}, we will neglect this contribution. The distribution amplitude of $n\bar{n}$ and $s\bar{s}$ contain same Lorentz structure with pion except for the difference decay constant and chiral parameters, we collect the relation \cite{Charng:2006zj}
\begin{eqnarray}
f_n&=&(1.07\pm0.02)f_{\pi}=139.1 \pm 2.6~\text{MeV}, \nonumber\\
f_s&=&(1.34\pm0.06)f_{\pi}=174.2 \pm 7.8~\text{MeV}.
\end{eqnarray}

The $S$-wave $\pi$-$\pi$ distribution amplitude takes the following form of~\cite{Diehl:1998dk, Muller:1994ses, Polyakov:1998ze, Hagler:2002nh, Meissner:2013hya}
\begin{eqnarray}
\Phi_{\pi\pi}^{S\text{-wave}}=\frac{1}{\sqrt{2N_c}}\bigg[p\hspace{-1.6truemm}/ \phi^{I=0}_{v\nu=-}(z,\zeta,\omega^2)+
\omega \phi^{I=0}_{s}(z,\zeta,\omega^2)+\omega(n\hspace{-2.0truemm}/ v\hspace{-1.8truemm}/ -1) \phi^{I=0}_{t\nu=+}(z,\zeta,\omega^2) \bigg],
\end{eqnarray}
where
\begin{eqnarray}
\phi^{I=0}_{v\nu=-}(z,\zeta,\omega^2)&=& \phi^0=\frac{9F_s(\omega^2)}{\sqrt{2N_c}}a_2^{I=0}z(1-z)(1-2z),\nonumber\\
\phi^{I=0}_{s}(z,\zeta,\omega^2)&=& \phi^s=\frac{F_s(\omega^2)}{2\sqrt{2N_c}}, \nonumber\\
\phi^{I=0}_{t\nu=+}(z,\zeta,\omega^2)&=& \phi^t=\frac{F_s(\omega^2)}{2\sqrt{2N_c}}(1-2z),
\end{eqnarray}
where the expression of scalar form factor and associated auxiliary functions can be found in~\cite{Wang:2015uea,Wang:2018fai}.



According to the typical Feynman diagrams as shown in Fig.~\ref{fig-fig1} and the quark currents for each decays, the decay amplitudes for considered quasi-two-body decays $B\to Pf_0(500)\to P\pi\pi$ are given as
\begin{eqnarray}
{\mathcal A}\big(B^+\to \pi^+ [\sigma\to]\pi^+\pi^-\big)&=&\frac{G_F}{2}\bigg\{V^*_{ub}V_{ud}\bigg[
\left(\frac{C_1}{3}+C_2\right)\left(F^{LL}_{T\sigma}+F^{LL}_{A\sigma}+F^{LL}_{AP}\right)+ C_1\left(M^{LL}_{T\sigma}+M^{LL}_{A\sigma}+M^{LL}_{AP}\right)
\nonumber \\
&+&C_2M^{LL}_{TP}\bigg]-V^*_{tb}V_{td}\bigg[\left(\frac{C_3}{3}+C_4+\frac{C_9}{3}+C_{10}\right) \left(F^{LL}_{T\sigma}+F^{LL}_{A\sigma}+F^{LL}_{AP}\right)
\nonumber \\
&+&\left(\frac{C_5}{3}+C_6+\frac{C_7}{3}+C_8\right) \left(F^{SP}_{T\sigma}+F^{SP}_{A\sigma}+F^{SP}_{AP}\right)
\nonumber \\
&+&\left(C_3+C_9\right) \left(M^{LL}_{T\sigma}+M^{LL}_{A\sigma}+M^{LL}_{AP}\right)
+\left(C_5+C_7\right) \left(M^{LR}_{T\sigma}+M^{LR}_{A\sigma}+M^{LR}_{AP}\right)
\nonumber \\
&+& \left(C_3+2C_4-\frac{C_9}{2}+\frac{C_{10}}{2}\right)M^{LL}_{TP}
\left(C_5-\frac{C_7}{2}\right)M^{LR}_{TP}
\nonumber \\
&+&\left(2C_6+\frac{C_8}{2}\right)M^{SP}_{TP}+
\left(\frac{C_5}{3}+C_6-\frac{C_7}{6}-\frac{C_8}{2}\right)F_{TP}^{SP}
\bigg]\bigg\}\;,
\\
{\mathcal A}\big(B^+\to K^+ [\sigma\to]\pi^+\pi^-\big)&=&\frac{G_F}{2}\bigg\{V^*_{ub}V_{us}\bigg[
\left(\frac{C_1}{3}+C_2\right)\left(F^{LL}_{T\sigma}+F^{LL}_{A\sigma}\right)+ C_1\left(M^{LL}_{T\sigma}+M^{LL}_{A\sigma}\right)+C_2M^{LL}_{TP}\bigg]
\nonumber \\
&-&V^*_{tb}V_{ts}\bigg[\left(\frac{C_3}{3}+C_4+\frac{C_9}{3}+C_{10}\right) \left(F^{LL}_{T\sigma}+F^{LL}_{A\sigma}\right)
\nonumber \\
&+&\left(\frac{C_5}{3}+C_6+\frac{C_7}{3}+C_8\right) \left(F^{SP}_{T\sigma}+F^{SP}_{A\sigma}\right)
+\left(C_3+C_9\right) \left(M^{LL}_{T\sigma}+M^{LL}_{A\sigma}\right)
\nonumber \\
&+&\left(C_5+C_7\right) \left(M^{LR}_{T\sigma}+M^{LR}_{A\sigma}\right)+ \left(2C_4+\frac{C_{10}}{2}\right)M^{LL}_{TP}
+\left(2C_6+\frac{C_8}{2}\right)M^{SP}_{TP}
\bigg]\bigg\}\;,
\\
{\mathcal A}\big(B^0\to \pi^0 [\sigma\to]\pi^+\pi^-\big)&=&\frac{G_F}{2\sqrt2}\bigg\{V^*_{ub}V_{ud}\bigg[
\left(C_1+\frac{C_2}{3}\right)\left(F^{LL}_{T\sigma}+F^{LL}_{A\sigma}+F^{LL}_{AP}\right)
\nonumber \\
&+& C_2\left(M^{LL}_{T\sigma}+M^{LL}_{A\sigma}+M^{LL}_{TP}+M^{LL}_{AP}\right)\bigg]
\nonumber \\
&-&V^*_{tb}V_{td}\bigg[\left(-\frac{C_3}{3}-C_4-\frac{3C_7}{2}-\frac{C_8}{2}+\frac{5C_9}{3}+C_{10}\right) \left(F^{LL}_{T\sigma}+F^{LL}_{A\sigma}+F^{LL}_{AP}\right)
\nonumber \\
&+&\left(-\frac{C_5}{3}-C_6+\frac{C_7}{6}+\frac{C_8}{2}\right) \left(F^{SP}_{T\sigma}+F^{SP}_{A\sigma}+F^{SP}_{TP}+F^{SP}_{AP}\right)
\nonumber \\
&+&\left(-C_3+\frac{C_9}{2}+\frac{3C_{10}}{2}\right) \left(M^{LL}_{T\sigma}+M^{LL}_{A\sigma}+M^{LL}_{AP}\right)
\nonumber \\
&+&\left(-C_3-2C_4+\frac{C_9}{2}-\frac{C_{10}}{2}\right) M_{TP}^{LL}+
\frac{3C_8}{2}\left(M^{SP}_{T\sigma}+M^{SP}_{A\sigma}+M^{SP}_{AP}\right)
\nonumber \\
&+&\left(-2C_6-\frac{C_8}{2}\right)M^{SP}_{TP}+\left(-C_5+\frac{C_7}{2}\right) \left(M^{LR}_{T\sigma}+M^{LR}_{A\sigma}+M^{LR}_{TP}+M^{LR}_{AP}\right)
\bigg]\bigg\}\;,
\\
{\mathcal A}\big(B^0\to K^0 [\sigma\to]\pi^+\pi^-\big)&=&\frac{G_F}{2}\bigg\{V^*_{ub}V_{us}\bigg[
C_2M^{LL}_{TP}\bigg]-V^*_{tb}V_{ts}\bigg[\left(\frac{C_3}{3}+C_4-\frac{C_9}{6}-\frac{C_{10}}{2}\right) \left(F^{LL}_{T\sigma}+F^{LL}_{A\sigma}\right)
\nonumber \\
&+&\left(\frac{C_5}{3}+C_6-\frac{C_7}{6}-\frac{C_8}{2}\right) \left(F^{SP}_{T\sigma}+F^{SP}_{A\sigma}\right)
+\left(C_3-\frac{C_9}{2}\right) \left(M^{LL}_{T\sigma}+M^{LL}_{A\sigma}\right)
\nonumber \\
&+&\left(C_5-\frac{C_7}{2}\right) \left(M^{LR}_{T\sigma}+M^{LR}_{A\sigma}\right)
+\left(2C_4+\frac{C_{10}}{2}\right)M^{LL}_{TP}+\left(2C_6+\frac{C_8}{2}\right)M^{SP}_{TP}
\bigg]\bigg\}\;,
\\
{\mathcal A}\big(B_s^0\to \bar{K}^0 [\sigma\to]\pi^+\pi^-\big)&=&\frac{G_F}{2}\bigg\{V^*_{ub}V_{ud}\bigg[
C_2M_{TP}^{LL}\bigg]-V^*_{tb}V_{td}\bigg[
\left(\frac{C_5}{3}+C_6-\frac{C_7}{6}-\frac{C_8}{2}\right)\left(F^{SP}_{TP}+F^{SP}_{AP}\right)
\nonumber \\
&+&\left(\frac{C_3}{3}+C_4-\frac{C_9}{6}-\frac{C_{10}}{2}\right)F_{AP}^{LL}
+\left(C_3+2C_4-\frac{C_9}{2}+\frac{C_{10}}{2}\right)M_{TP}^{LL}
\nonumber \\
&+&\left(C_5-\frac{C_7}{2}\right)M_{TP}^{LR}+
\left(2C_6-\frac{C_8}{2}\right)M_{TP}^{SP}
+\left(C_3-\frac{C_9}{2}\right)M_{AP}^{LL}
\nonumber \\
&+&\left(C_5-\frac{C_7}{2}\right)M_{AP}^{LR}
\bigg]\bigg\}\;,
\\
{\mathcal A}\big(B^0\to \eta_q [\sigma\to]\pi^+\pi^-\big)&=&\frac{G_F}{2\sqrt2}\bigg\{V^*_{ub}V_{ud}\bigg[
\left(C_1+\frac{C_2}{3}\right)\left(F_{T\sigma}^{LL}+F_{A\sigma}^{LL}+F_{AP}^{LL}\right)
\nonumber\\
&+&C_2\left(M_{T\sigma}^{LL}+M_{A\sigma}^{LL}+M_{TP}^{LL}+M_{AP}^{LL}\right)\bigg]
\nonumber \\
&-&V^*_{tb}V_{td}\bigg[
\left(C_5+\frac{C_6}{3}-\frac{C_7}{6}-\frac{C_8}{2}\right)\left(F^{SP}_{T\sigma}+F^{SP}_{TP}\right)
+\left(C_6-\frac{C_8}{2}\right)\left(M_{T\sigma}^{SP}+M_{TP}^{SP}\right)
\nonumber \\
&+&\left(\frac{7C_3}{3}+\frac{7C_4}{3}-2C_5-\frac{2C_6}{3}-\frac{C_7}{2}-\frac{C_8}{6}+\frac{C_9}{3} -\frac{C_{10}}{3}\right)\left(F_{T\sigma}^{LL}+F_{A\sigma}^{LL}+F_{AP}^{LL}\right)
\nonumber \\
&+&\left(C_3+2C_4-\frac{C_9}{2}+\frac{C_{10}}{2}\right)\left(M_{T\sigma}^{LL}+M_{TP}^{LL}\right)
\nonumber \\
&+&\left(C_5+C_6-\frac{C_7}{2}+C_8\right)\left(M_{T\sigma}^{LR}+M_{TP}^{LR}\right)
\nonumber \\
&+&\left(\frac{C_5}{3}+C_6-\frac{C_7}{6}-\frac{C_8}{2}\right)\left(F_{A\sigma}^{SP}+M_{AP}^{SP}\right)+
\left(C_4+C_{10}\right)\left(M_{A\sigma}^{LL}+M_{AP}^{LL}\right)
\nonumber \\
&+&\left(C_6+C_8\right)\left(M_{A\sigma}^{SP}+M_{AP}^{SP}\right)
\bigg]\bigg\}\;,
\\
{\mathcal A}\big(B^0\to \eta_s [\sigma\to]\pi^+\pi^-\big)&=&\frac{G_F}{2}\bigg\{
-V^*_{tb}V_{td}\bigg[\left(C_4-\frac{C_{10}}{2}\right)M_{T\sigma}^{LL}+\left(C_6-\frac{C_8}{2}\right)M_{T\sigma}^{SP}
\nonumber \\
&+&\left(C_3+\frac{C_4}{3}-C_5-\frac{C_6}{3}+\frac{C_7}{2}+\frac{C_8}{6}-\frac{C_9}{2} -\frac{C_{10}}{6}\right)F_{T\sigma}^{LL}
\bigg]\bigg\}\;,
\\
{\mathcal A}\big(B^0\to \eta[\sigma\to]\pi^+\pi^-\big)&=&{\mathcal A}\big(B^0\to \eta_q[\sigma\to]\pi^+\pi^-\big) \cos\phi-{\mathcal A}\big(B^0\to \eta_s[\sigma\to]\pi^+\pi^-\big)\sin\phi \;,
\\
{\mathcal A}\big(B^0\to \eta^{\prime}[\sigma\to]\pi^+\pi^-\big)&=&{\mathcal A}\big(B^0\to \eta_q[\sigma\to]\pi^+\pi^-\big)\sin\phi+ {\mathcal A}\big(B^0\to \eta_s[\sigma\to]\pi^+\pi^-\big)\cos\phi \;,
\\
{\mathcal A}\big(B_s^0\to \eta_q [\sigma\to]\pi^+\pi^-\big)&=&\frac{G_F}{2\sqrt2}\bigg\{V^*_{ub}V_{us}\bigg[
\left(C_1+\frac{C_2}{3}\right)\left(F_{A\sigma}^{LL}+F_{AP}^{LL}\right)
+C_2\left(M_{A\sigma}^{LL}+M_{AP}^{LL}\right)\bigg]
\nonumber \\
&-&V^*_{tb}V_{ts}\bigg[
\left(2C_3+\frac{2C_4}{3}-2C_5-\frac{2C_6}{3}-\frac{C_7}{2}-\frac{C_8}{6}+\frac{C_9}{2} +\frac{C_{10}}{6}\right)\left(F_{A\sigma}^{LL}+F_{AP}^{LL}\right)
\nonumber \\
&+&\left(2C_4+\frac{C_{10}}{2}\right)\left(M_{A\sigma}^{LL}+M_{AP}^{LL}\right)+
\left(2C_6+\frac{C_8}{2}\right)\left(M_{A\sigma}^{SP}+M_{AP}^{SP}\right)
\bigg]\bigg\}\;,
\\
{\mathcal A}\big(B_s^0\to \eta_s [\sigma\to]\pi^+\pi^-\big)&=&\frac{G_F}{2}\bigg\{V^*_{ub}V_{us}\bigg[
C_2M_{TP}^{LL}\bigg]
-V^*_{tb}V_{ts}\bigg[\left(C_4+C_{10}\right)M_{TP}^{LL}+\left(C_6+C_8\right)M_{TP}^{SP}
\bigg]\bigg\}\;,
\\
{\mathcal A}\big(B_s^0\to \eta[\sigma\to]\pi^+\pi^-\big)&=&{\mathcal A}\big(B_s^0\to \eta_q[\sigma\to]\pi^+\pi^-\big) \cos\phi-{\mathcal A}\big(B_s^0\to \eta_s[\sigma\to]\pi^+\pi^-\big)\sin\phi \;,
\\
{\mathcal A}\big(B_s^0\to \eta^{\prime}[\sigma\to]\pi^+\pi^-\big)&=&{\mathcal A}\big(B_s^0\to \eta_q[\sigma\to]\pi^+\pi^-\big)\sin\phi+ {\mathcal A}\big(B_s^0\to \eta_s[\sigma\to]\pi^+\pi^-\big)\cos\phi \;.
\end{eqnarray}
where $G_F$ is the Fermi constant, $V_{ij}$ is the CKM matrix element, and the
combinations of the Wilson coefficients $a_{1,2}$ are defined as $a_1=C_1/3+C_2$ and $a_2=C_2/{3}+C_1$.
The expressions of individual amplitudes $F^{LL}_{T\sigma}$, $F^{SP}_{T\sigma}$, $F^{LL}_{A\sigma}$, $F^{SP}_{A\sigma}$, $M^{LL}_{T\sigma}$,~$M^{LR}_{T\sigma}$,~$M^{SP}_{T\sigma}$, $M^{LL}_{A\sigma}$,~$M^{LR}_{A\sigma}$,~$M^{SP}_{A\sigma}$, $F^{SP}_{TP}$,~$M^{LL}_{TP}$,~$M^{LR}_{TP}$,~$M^{SP}_{TP}$, $F^{LL}_{AP}$,~$F^{SP}_{AP}$, $M^{LL}_{AP}$,~$M^{LR}_{AP}$ and $M^{SP}_{AP}$
from different subdiagrams in Fig.~\ref{fig-fig1}, which are collected in the Appendix~\ref{appendix A}.

Differential branching ratio of $B \to P f_0(500) \to P \pi \pi$ can be written as
\begin{eqnarray}\label{Br}
\frac{d\mathcal{B}}{d\eta}=\tau_{B} \frac{\vert \vec{p}_1\vert \vert \vec{p}\vert B^2_0 C^2 }{32 \pi^3  m_B m^2_0 }\overline{|{\cal A}|^2} \;,\label{eqn-diff-bra}
\end{eqnarray}
where $B_0$ is proportional to quark condensate, in $\pi\pi$ scalar system, it can be parameterized as $B_0 \simeq m_{\pi}^2/(m_u+m_d)$. Following the definition from Ref.~\cite{Wang:2018fai}, the constant $C$ takes the form of $C={g_{\sigma\pi\pi} \bar{f}_{\sigma}} /(\sqrt{2}B_0m_0)$. Meanwhile, $\tau_{B}$ is the mean lifetime of $B$ meson, $\vert\vec{p}_1\vert$
and $\vert\vec{p}\vert$ are the three momenta of $f_0(500)$ resonance and light pseudoscalar meson respectively in the center-of-mass frame of $\pi$-$\pi$, and can be written as
\begin{eqnarray}
\vert \overrightarrow{p_1}\vert=\frac{\sqrt{\lambda(\omega^2,m_{\pi}^2,m^2_{\pi})}}{2\omega}, \quad
\vert \overrightarrow{p}\vert=\frac{\sqrt{\lambda(m_B^2,m_{P}^2,\omega^2)}}{2\omega},
\end{eqnarray}
with $m_{B}$, $m_P$ and $m_{\pi}$ are the masses of the $B$, light pseudoscalar and $\pi$ mesons respectively, and the K{\"a}ll{\'e}n function $\lambda(a,b,c)=a^2+b^2+c^2-2(ab+ac+bc)$.

\section{RESULTS}\label{section3}
\begin{table}[h]
\centering
\tabcolsep=0.35cm
\renewcommand{\arraystretch}{2.0}
\begin{tabular}{| c | c c c | c | c c c |}
    \hline\hline
 Parameter  & Value    & Unit       & Reference  & Parameter  & Value    & Unit       & Reference \\
    \hline\hline
   $m_{B^{\pm}}$ &  5.279 & GeV  & \cite{ParticleDataGroup:2022pth} &
   $\tau_{B^{\pm}}$ &  1.638 & ps  & \cite{ParticleDataGroup:2022pth} \\
   $m_{B^0}$ &  5.280 & GeV  & \cite{ParticleDataGroup:2022pth} &
   $\tau_{B^0}$ &  1.519 & ps  & \cite{ParticleDataGroup:2022pth} \\
   $m_{B_s}$ &  5.367 & GeV  & \cite{ParticleDataGroup:2022pth} &
   $\tau_{B_s}$ &  1.527 & ps  & \cite{ParticleDataGroup:2022pth} \\
    $f_B|_{N_f=2+1+1}$ &  190.0 & MeV  &\cite{FlavourLatticeAveragingGroupFLAG:2021npn}&
   $f_{B_s}|_{N_f=2+1+1}$ &  230.3 & MeV  &\cite{FlavourLatticeAveragingGroupFLAG:2021npn}  \\
   \hline\hline
   $m_{\pi^{\pm}}$ & 0.140  & GeV  & \cite{ParticleDataGroup:2022pth} &
   $m_{\pi^0}$ &  0.135 & GeV  & \cite{ParticleDataGroup:2022pth} \\
   $m_{K^{\pm}}$ &  0.494 & GeV  & \cite{ParticleDataGroup:2022pth} &
    $m_{K^0}$ &  0.498 & GeV  & \cite{ParticleDataGroup:2022pth} \\
    $m_{\eta}$ &  0.548 & GeV  & \cite{ParticleDataGroup:2022pth} &
   $m_{\eta^{\prime}}$ &  0.958  & GeV  & \cite{ParticleDataGroup:2022pth} \\
   $f_{\pi}$ &  0.130 & GeV  & \cite{ParticleDataGroup:2022pth} &
   $f_K$ &  0.156  & GeV  & \cite{ParticleDataGroup:2022pth} \\
   \hline\hline
    $\lambda$ &  0.2250  & ~ &\cite{ParticleDataGroup:2022pth} &
     $\bar{\rho}$ &  0.159 & ~ & \cite{ParticleDataGroup:2022pth}  \\
     $A$ &  0.826 & ~ & \cite{ParticleDataGroup:2022pth} &
     $\bar{\eta}$ &  0.348 & ~ &  \cite{ParticleDataGroup:2022pth}\\
     \hline\hline
\end{tabular}
\caption{Numerical values of the theory input parameters employed in the PQCD predictions of the quasi-two-body decays $\to P f(500)\to P\pi\pi$ as well as the subsequent phenomenological
analysis for the quasi-two-body decay observables.}
\label{Inputparameters}
\end{table}

For the numerical calculation, we adopt QCD scale at $\mu=0.25$~GeV in modified minimal subtraction scheme. The decay constant of $B^{0,\pm}$, $B_s$ and light pseudoscalar mesons come from FLAG working group's result \cite{FlavourLatticeAveragingGroupFLAG:2021npn}. The masses and mean life times of $B^{0,\pm}$ and $B_s$ mesons, the masses of light pseudoscalar mesons and Wolfenstein parameters are all come from recent updated \textit{Review of Particle Physics} \cite{ParticleDataGroup:2022pth}. We summarize explicitly the numerical values of the necessary standard model inputs and the hadronic parameters in Table \ref{Inputparameters}. Here the subscript $N_f$ in decay constant $f_{B_{(s)}}$ represents the number of dynamical quark flavor in lattice simulation. $N_f=2+1+1$ is for $m_u=m_d<m_s<m_c$ with four flavors dynamical quarks. The result for $N_f=2 + 1 + 1$ is considered to be the most realistic one in comparing with $N_f = 2$ and $N_f = 2 + 1$, which can be found by the Flavor Lattice Averaging Group in detail~\cite{FlavourLatticeAveragingGroupFLAG:2021npn}.

By using the formula of differential branching fraction in Eq.~(\ref{eqn-diff-bra}) and explicit decay amplitude in Appendix~\ref{appendix A}, we obtain the PQCD predictions of $CP$ averaged branching fractions and direct $CP$ violations in Table~\ref{BRCP} for quasi-two-body $B\to P f(500)  \to P \pi\pi$ processes. The first uncertainty comes from shape parameter $\omega_B$ in the $B$ meson distribution amplitude, we vary it value at 10\% magnitude, that is $\omega_{B^{0,\pm}}=0.40 \pm 0.04$ GeV and $\omega_{B_s}=0.5 \pm 0.05$ GeV for $B^{0,\pm}$ and $B_s$ mesons respectively. The second uncertainty comes from the Gegenbauer moment $a_2^{I=0}=0.20 \pm 0.20$
in the $S$-wave $\pi$-$\pi$ distribution amplitude. We ignore the uncertainties of parameters in distribution amplitude of light pseudoscalar mesons and Wolfenstein parameters since this uncertainties are considered to be small. It is interesting to see that the $\omega_B$ produce the largest theoretical uncertainty in PQCD predictions of quasi-two-body $B$ meson decays~\cite{Wang:2016rlo, Ma:2017idu, Ma:2017kec, Rui:2017hks, Li:2018lbd, Wang:2018dfq, Ma:2019qlm, Cui:2019khu, Wang:2020saq, Fan:2020gvr, Chai:2021pyp}, but in this study and previous quasi-two-body $B$ meson decays involving $S$-wave $\pi$-$\pi$ contributions~\cite{Wang:2015uea, Ma:2017idu, Liu:2019ymi, Xing:2019xti}, the Gegenbauer moment $a_2^{I=0}$ plays more important role.
\begin{table}[t]
\centering
\tabcolsep=0.3cm
\renewcommand{\arraystretch}{2.2}
\begin{tabular}{ccc}
    \hline \hline
 Decay modes  ~~~~&~~ ~ ~~&~~ Quasi-two-body results \\
 \hline
  $B^+\to \pi^+ [\sigma\to]\pi^+\pi^-$ & $\mathcal{B}(10^{-6})$ & $1.78 \pm 0.41 \;(\omega_B) \pm 0.51\;(a_2)$   \\
  $ $ & $\mathcal{A}_{CP}(\%)$ & $29.8\pm 11.1\;(\omega_B) \pm 13.0\;(a_2)$  \\
  \hline
  $B^+\to K^+ [\sigma\to]\pi^+\pi^-$ & $\mathcal{B}(10^{-7})$ & $8.14\pm 1.26 \;(\omega_B) \pm 1.10\;(a_2)$  \\
  $ $ & $\mathcal{A}_{CP}(\%)$ & $-63.5\pm 8.2 \;(\omega_B) \pm 11.3\;(a_2)$  \\
  \hline
  $B^0\to \pi^0 [\sigma\to]\pi^+\pi^-$ & $\mathcal{B}(10^{-7})$ & $2.02 \pm 0.55 \;(\omega_B) \pm 1.23\;(a_2)$ \\
  $ $ & $\mathcal{A}_{CP}(\%)$ & $-64.6 \pm 21.8 \;(\omega_B) \pm 38.7\;(a_2)$  \\
  \hline
  $B^0\to K^0 [\sigma\to]\pi^+\pi^-$ & $\mathcal{B}(10^{-7})$ & $3.90 \pm 0.26 \;(\omega_B) \pm 2.68\;(a_2)$   \\
  $ $ & $\mathcal{A}_{CP}(\%)$ & $1.31\pm 18.0 \;(\omega_B) \pm 10.6\;(a_2)$  \\
  \hline
  $B_s^0\to \bar{K}^0 [\sigma\to]\pi^+\pi^-$ & $\mathcal{B}(10^{-7})$ & $1.18\pm 0.22 \;(\omega_B) \pm 0.72\;(a_2)$ \\
  $ $ & $\mathcal{A}_{CP}(\%)$ & $2.26\pm 8.61 \;(\omega_B) \pm 8.13\;(a_2)$  \\
  \hline
  $B^0\to \eta[\sigma\to]\pi^+\pi^-$ & $\mathcal{B}(10^{-8})$ & $5.11\pm 1.11 \;(\omega_B) \pm 4.60\;(a_2)$   \\
  $ $ & $\mathcal{A}_{CP}(\%)$ & $-88.9\pm 1.3 \;(\omega_B) \pm 47.24\;(a_2)$  \\
  \hline
  $B^0\to \eta^{\prime}[\sigma\to]\pi^+\pi^-$ & $\mathcal{B}(10^{-8})$ & $2.29\pm0.61\;(\omega_B) \pm  1.59 \;(a_2)$  \\
  $ $ & $\mathcal{A}_{CP}(\%)$ & $-42.3\pm 13.4 \;(\omega_B) \pm 5.4\;(a_2)$  \\
  \hline
  $B_s^0\to \eta[\sigma\to]\pi^+\pi^-$ & $\mathcal{B}(10^{-9})$ & $2.00\pm 1.04 \;(\omega_B) \pm 1.25\;(a_2)$  \\
  $ $ & $\mathcal{A}_{CP}(\%)$ & $15.2\pm 35.1 \;(\omega_B) \pm 17.1\;(a_2)$  \\
  \hline
  $B_s^0\to \eta^{\prime}[\sigma\to]\pi^+\pi^-$ & $\mathcal{B}(10^{-8})$ & $2.33\pm 0.72 \;(\omega_B) \pm 1.05\;(a_2)$  \\
  $ $ & $\mathcal{A}_{CP}(\%)$ & $10.4\pm 3.9 \;(\omega_B) \pm 6.8\;(a_2)$  \\
  \hline\hline
\end{tabular}
\caption{PQCD predictions of $CP$-averaged branching fraction and direct
$CP$ violation for the quasi-two-body decays
$B\to P f(500)  \to P \pi\pi$. }
\label{BRCP}
\end{table}

From the numerical results as listed in Table~\ref{BRCP}, we have the following comments:
\begin{itemize}
  \item In the $B\to PR\to P\pi\pi$ decays, we can extract the two-body branching fractions $\mathcal{B}(B\to PR)$ by using the relation under the quasi-two-body approximation
  \begin{equation}\label{NWA}
     \mathcal{B}(B\to PR\to P\pi\pi)=\mathcal{B}(B\to PR) \cdot \mathcal{B}(R\to \pi\pi)\;.
  \end{equation}
      Combined with results listed in Table~\ref{BRCP}, one can obtain the related two-body branching fractions for the two-body decays $B\to P f_0(500)$. It is interesting to see that the finite width effect are prominent in $B\to P f_0(500)$ decay~\cite{Cheng:2020iwk}, which means the extraction of two-body results by using quasi-two-body approximation will be greatly underestimated.

  \item The predicted $CP$-averaged branching fractions for the considered decay processes are in the range of $10^{-9}-10^{-6}$. For the $B^+ \to \pi^+ f_0(500) \to \pi^+ \pi^+ \pi^-$ decay, our prediction $(1.78 \pm 0.41\pm 0.51) \times 10^{-6}$ is agree with Ref.~\cite{Cheng:2020iwk} from the QCDF prediction by considering the finite width effect, both of two predictions satisfy the upper limit $4.1\times 10^{-6}$ from $BABAR$ measurement~\cite{BaBar:2005jqu}.  However, most resent LHCb measurement~\cite{LHCb:2019sus,LHCb:2019jta} for the  $B^+ \to \pi^+ f_0(500) \to \pi^+ \pi^+ \pi^-$ decay in the context of isobar model yields $(3.83\pm 0.84) \times 10^{-6}$, the PQCD predicted branching fraction is smaller than the LHCb measurement by a factor of about 2. To have a clear look at the above comparison, we present the predictions in Table~\ref{BRCP-comparison}. Furthermore, we show the curve of $\pi$-$\pi$ invariant mass dependent differential branching fraction for quasi-two-body decay $B^+ \to \pi^+ f_0(500) \to \pi^+ \pi^+ \pi^-$ in Fig.~\ref{fig-BR-wdep}. It is found that the main portion of branching fraction for $B^+ \to \pi^+ f_0(500) \to \pi^+ \pi^+ \pi^-$ received from the region of $[2\, m_{\pi}, 1\, {\rm GeV}]$, the contributions from the  $m_{\pi\pi}>2.0$GeV is evaluated about $2.4\%$ and can be neglected safely.

\begin{table}[t]
\centering
\tabcolsep=0.3cm
\renewcommand{\arraystretch}{1.5}
\begin{tabular}{lll}
\hline \hline
References ~~~~&~~ $\mathcal{B}(10^{-6})$ ~~&~~ $\mathcal{A}_{CP}(\%)$ \\ \hline
This Work & $1.78 \pm 0.41\pm 0.51$  & $29.8\pm 11.1 \pm 13.0$          \\
QCDF~\cite{Cheng:2020iwk}  & $1.65^{+0.42}_{-0.37}$  & $14.7\pm0.1$     \\
BABAR~\cite{BaBar:2005jqu} & $<4.1$                  & -                \\
LHCb~\cite{LHCb:2019sus,LHCb:2019jta}      & $3.83\pm 0.84$   & $14.9^{+0.5}_{-0.6}$             \\
\hline\hline
\end{tabular}
\caption{Predictions about $CP$-averaged branching fraction and direct
$CP$ violation for the quasi-two-body decays $B^+ \to \pi^+ f_0(500) \to \pi^+ \pi^+ \pi^-$ decay. Meanwhile, we also listed the QCDF, BABAR, LHCb results as a comparison.}
\label{BRCP-comparison}
\end{table}

\begin{figure}[t]
\centerline{\epsfxsize=9cm\epsffile{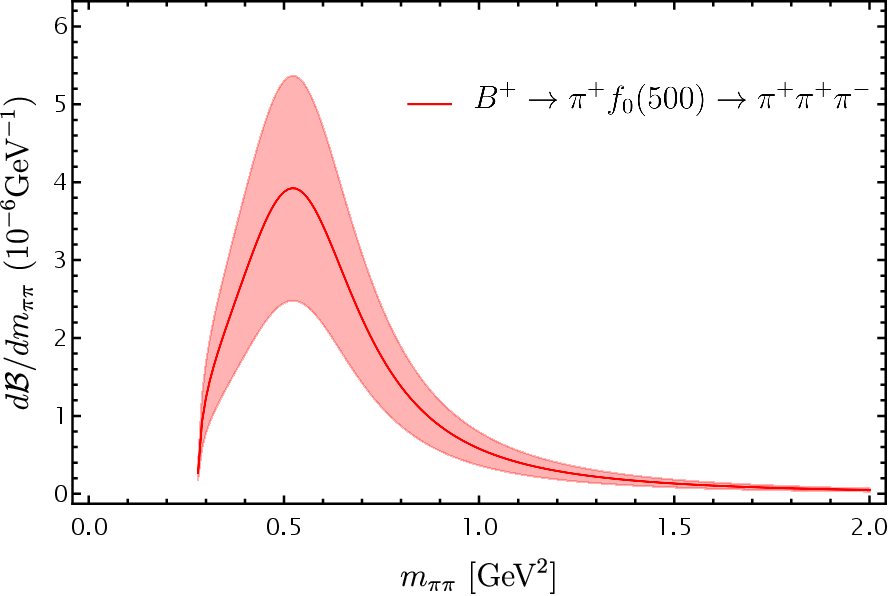}}
\vspace{0.3cm}
\caption{The $\pi\pi$ invariant mass-dependent differential branching fraction for quasi-two-body decay $B^+ \to \pi^+ f_0(500) \to \pi^+ \pi^+ \pi^-$, the uncertainties generated from $\omega_B$ and $a_2$ are shown by the shaded band.}
\label{fig-BR-wdep}
\end{figure}

  \item For the quasi-two-body decay $B^+ \to \pi^+ f_0(500) \to \pi^+ \pi^+ \pi^-$, our prediction of $CP$ violation reads $(29.8\pm 11.1\pm 13.0)\%$, which is much greater than that $14.9^{+0.5}_{-0.6} \%$ from LHCb measurements~\cite{LHCb:2019sus,LHCb:2019jta} in the context of isobar model for the $S$-wave $\pi$-$\pi$, fortunately, this result is still in the range of our prediction by considering the uncertainty from two nonperturbative parameters, which can also be seen in Table~\ref{BRCP-comparison}. We also mention that QCDF prediction~\cite{Cheng:2020iwk} is in excellent agreement with LHCb result. Besides, LHCb collaboration found that the interference between the $S$- and $P$-waves can also generate $CP$ violation effect, unfortunately, $CP$ violation can only produced from interference of tree and penguin diagrams in the state of art of PQCD calculations, the $S$- and $P$-waves cannot generate extra strong phase, the predictions of $CP$ violation from interference of $S$- and $P$-waves is still absent and very challenging in PQCD. For sake of illustrate $\pi$-$\pi$ invariant mass dependent $CP$ asymmetry, we show the $\pi$-$\pi$ invariant mass-dependent $CP$ asymmetry for quasi-two-body decay $B^+ \to \pi^+ f_0(500) \to \pi^+ \pi^+ \pi^-$ as an example, it is interesting to see that the direct $CP$ asymmetry is decrease as the $\pi$-$\pi$ invariant mass increases.

  \begin{figure}[t]
\centerline{\epsfxsize=10cm \epsffile{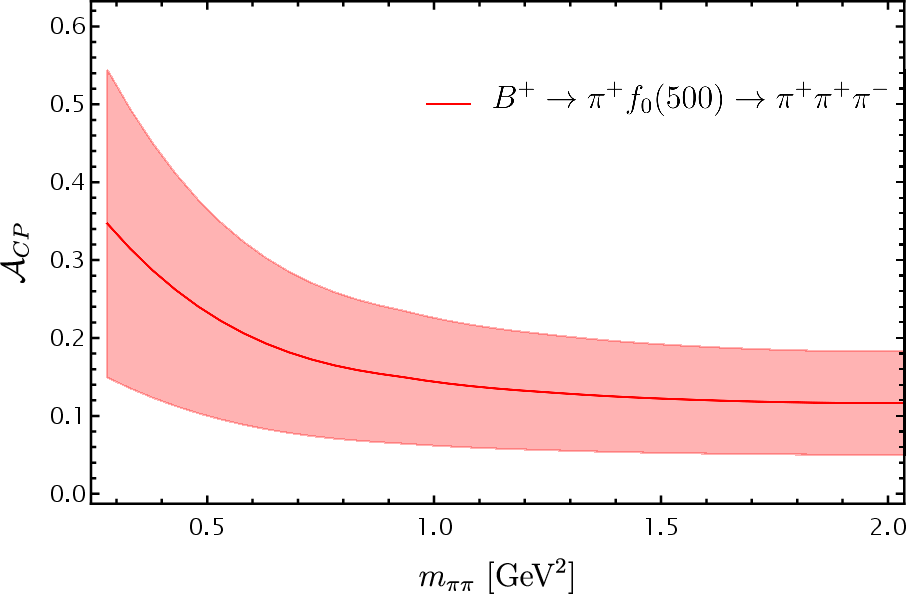}}
\vspace{0.3cm}
\caption{The $\pi\pi$ invariant mass-dependent $CP$ asymmetry for quasi-two-body decay $B^+ \to \pi^+ f_0(500) \to \pi^+ \pi^+ \pi^-$, the uncertainties generated from $\omega_B$ and $a_2$ are shown by the shaded band.}
\label{fig-Acp-wdep}
\end{figure}

  \item $f_0(500)$ was often parameterized by Bugg model~\cite{Bugg:2006gc} in partial wave analysis and it was applied in LHCb measurements~\cite{LHCb:2014xpr,LHCb:2015klp}. However, $S$-wave $\pi$-$\pi$ are considered to be broad, overlapping resonances, $K$-matrix model was applied for parametrization of $S$-wave components as a alternative scheme~\cite{LHCb:2019sus,LHCb:2019jta}. The rigorous theoretical calculation for nonresonance contribution in the context of PQCD framework is still absent~\cite{Wang:2020saq}, comparison between experiment measurements and theoretical predictions is still challenging. More attempts can be make in future study to parameterize the nonresonance contribution for sake of giving a more reliable result.

  \item It is also worth mentioning that the Gegenbauer moment $a_2$ plays very different roles in different decay mode. First, the annihilation diagram will generate very large uncertainty from Gegenbauer moment $a_2$. The flavor changing charged current mode such as $B^+\to K^+[\sigma\to]\pi^+\pi^-$ decay processes are transition diagram dominated, which can be seen in subdiagram (a) of Fig.~1. Meanwhile, the flavor changing neutral current mode such as $B^0\to K^0[\sigma\to]\pi^+\pi^-$ are annihilation diagram dominated, which can be seen in subdiagram (b) of Fig.~1. Taking decay process $B^+\to K^+[\sigma\to]\pi^+\pi^-$ and $B^0\to K^0[\sigma\to]\pi^+\pi^-$ as example, we can make a comparison for contributions from different diagrams. It is found that annihilation diagram will give about $72.3\%$ contribution for branching fraction in decay mode $B^0\to K^0[\sigma\to]\pi^+\pi^-$, therefore, the uncertainty generated by $a_2$ will very large. On the contrary, annihilation diagram will give just about $35.0\%$ contribution for branching fraction in decay mode $B^+\to K^+[\sigma\to]\pi^+\pi^-$. Thus, the large uncertainty from Gegenbauer moment $a_2$ will be amortized and become much lower than that of annihilation dominated mode. Subsequently, the flavor changing charged current mode are tree dominated, while the neutral mode are penguin dominated. The very complicated complex interference from CKM matrix, Wilson coefficients and form factors will enlarge or reduce the uncertainty caused by $a_2$, which encoding the very different behaviors in different mode.
\end{itemize}

\section{CONCLUSION}\label{section4}

We have studied the quasi-two-body decays $B \to P f_0(500) \to P \pi^+ \pi^-$ in the framework of PQCD factorization approach by using the scalar form factor $F_{\pi\pi}(\omega^2)$ as the nonperturbative input. Here the bachelor particle $P$ denotes $\pi$, $K$, $\eta$ and $\eta^{\prime}$. The $CP$ averaged branching fractions and the $CP$ asymmetries for the considered quasi-two-body decay modes have been calculated by using the quasi-two-body approximation. It is found that the branching fractions are within the range of $10^{-9}-10^{-6}$. Our predicted branching fraction for $B^+ \to \pi^+ f_0(500) \to \pi^+ \pi^+ \pi^-$ agrees with the upper limit issued by the $BABAR$ collaboration but is smaller than the LHCb measurement by a factor of $\sim 2$. In Figs.~\ref{fig-BR-wdep} and \ref{fig-Acp-wdep}, we have shown the differential branching fraction over the $\pi$-$\pi$ invariant mass and the direct $CP$ violation for the channel $B^+ \to \pi^+ f_0(500) \to \pi^+ \pi^+ \pi^-$. We hope all the predictions could be tested by further experiments, it may deeper our understanding for the $B$ meson decay mechanism in the context of PQCD factorization approach, and help us to probe the inner structure of $f_0(500)$.

\acknowledgments
This work is supported by National Natural Science Foundation of China under Contract No. 12275036, No. 12347101, No. 12265010, the Natural Science Foundation of Chongqing under Contract No.cstc2021jcyjmsxmX0681, the Science
and Technology Research Program of Chongqing Municipal Education Commission under Contract
No.KJQN202001541 and the Research Foundation of Chongqing University of Science and Technology under the project No.ckrc20231220.

\appendix

\section{Decay Amplitudes}\label{appendix A}
The factorization formulas for decay amplitudes from Fig.~\ref{fig-fig1} are collected below
\begin{eqnarray}
F^{LL}_{T\sigma} &=& 8\pi C_F m^4_B f_P\int dx_B dz\int b_B db_B b db \phi_B(x_B,b_B)(1-\eta)
\big\{\big[\sqrt{\eta}(1-2z)(\phi^s+\phi^t)+(1+z)\phi^0 \big]
\nonumber\\
&\times&E_{1ab}(t_{1a})h_{1a}(x_B,z,b_B,b)S_t(z)+\sqrt{\eta}\left(2\phi^s-\sqrt\eta\phi^0 \right)E_{1ab}(t_{1b})h_{1b}(x_B,z,b_B,b)S_t(x_B) \big\}\;,
\\
F^{SP}_{T\sigma} &=&-16\pi C_F m^4_B r_0 f_P \int dx_B dz\int b_B db_B b db \phi_B(x_B,b_B)
\big\{\left[\sqrt{\eta}(2+z)\phi^s-\sqrt{\eta}z\phi^t+(1+\eta(1-2z))\phi^0 \right]
\nonumber\\
&\times& E_{1ab}(t_{1a})h_{1a}(x_B,z,b_B,b)S_t(z)+\left[2\sqrt{\eta}(1-x_B+\eta)\phi^s+(x_B-2\eta)\phi^0 \right]
\nonumber\\
&\times& E_{1ab}(t_{1b})h_{1b}(x_B,z,b_B,b)S_t(x_B)\big\}\;,
\\
M^{LL}_{T\sigma} &=& 32\pi C_F m^4_B/\sqrt{2N_c} \int dx_B dz dx_3\int b_B db_B b_3 db_3
\phi_B(x_B,b_B)\phi^A(1-\eta)
\nonumber\\
&\times&\big\{\left[\sqrt{\eta}z(\phi^t-\phi^s)+((1-\eta)(1-x_3)-x_B+z\eta)\phi^0 \right]
E_{1cd}(t_{1c})h_{1c}(x_B,z,x_3,b_B,b_3)
\nonumber\\
&+&\left[z(\sqrt{\eta}(\phi^s+\phi^t)-\phi^0 )-(x_3(1-\eta)-x_B)\phi^0 \right]E_{1cd}(t_{1d})
h_{1d}(x_B,z,x_3,b_B,b_3) \big\}\;,
\\
M^{LR}_{T\sigma} &=& -32\pi C_F r_0 m^4_B/\sqrt{2N_c}\int dx_B dz dx_3\int b_B db_B b_3 db_3
\phi_B(x_B,b_B)\big\{\big[\sqrt{\eta}z(\phi^P-\phi^T)(\phi^s+\phi^t)
\nonumber\\
&+&\sqrt{\eta}((1-x_3)(1-\eta)-x_B)(\phi^P+\phi^T)(\phi^s-\phi^t)
+((1-x_3)(1-\eta)-x_B)(\phi^P+\phi^T)\phi^0
\nonumber\\
&+&\eta z(\phi^P-\phi^T)\phi^0 \big] E_{1cd}(t_{1c})h_{1c}(x_B,z,x_3,b_B,b_3)
+\big[-\sqrt{\eta}z(\phi^P+\phi^T)(\sqrt{\eta}\phi^0+(\phi^t+\phi^s))
\nonumber\\
&+&(x_B-x_3(1-\eta))(\phi^P-\phi^T)
(\sqrt{\eta}(\phi^s-\phi^t)+\phi^0)\big] E_{1cd}(t_{1d})h_{1d}(x_B,z,x_3,b_B,b_3)\big\}\;,
\\
M^{SP}_{T\sigma} &=& 32\pi C_F m^4_B/\sqrt{2N_c} \int dx_B dz dx_3\int b_B db_B b_3 db_3
\phi_B(x_B,b_B)\phi^A(\eta-1)
\nonumber\\
&\times&\big\{\left[\sqrt{\eta}z(\phi^t+\phi^s)+((\eta-1)(1-x_3)+x_B-z\eta)\phi^0 \right]
E_{1cd}(t_{1c})h_{1c}(x_B,z,x_3,b_B,b_3)
\nonumber\\
&+&\left[z(\sqrt{\eta}(\phi^s-\phi^t)-\eta\phi^0 )-(x_3(1-\eta)-x_B)\phi^0 \right]E_{1cd}(t_{1d})
h_{1d}(x_B,z,x_3,b_B,b_3) \big\}\;,
\\
F^{LL}_{A\sigma} &=& 8\pi C_F m^4_B f_B\int dz dx_3\int b db b_3 db_3 \big\{\left[2r_0\sqrt{\eta}\phi^P
((2-z)\phi^s+z\phi^t)-(1-\eta)(1-z)\phi^A\phi^0\right] E_{1ef}(t_{1e})
\nonumber\\
&\times& h_{1e}(z,x_3,b,b_3)S_t(z) +\big[2r_0\sqrt{\eta}[(1-x_3)(1-\eta)\phi^T
-(1+x_3+(1-x_3)\eta)\phi^P]\phi^s
\nonumber\\
&+&(x_3(1-\eta)+\eta)(1-\eta)\phi^A\phi^0 \big] E_{1ef}(t_{1f})h_{1f}(z,x_3,b,b_3)S_t(x_3) \big\}\;,
\\
F^{SP}_{A\sigma} &=&  16\pi C_F m^4_B f_B \int dz dx_3\int b db b_3 db_3
\big\{\left[\sqrt{\eta}(1-\eta)(1-z)\phi^A(\phi^s+\phi^t)-2r_0(1+(1-z)\eta)\phi^P\phi^0 \right]
\nonumber\\
&\times&E_{1ef}(t_{1e})  h_{1e}(z,x_3,b,b_3)S_t(z) +\left[2\sqrt{\eta}(1-\eta)\phi^A\phi^s - r_0\left(2\eta \phi^P
+x_3(1-\eta)(\phi^P-\phi^T)\right)\phi^0\right]
\nonumber\\
&\times&E_{1ef}(t_{1f})h_{1f}(z,x_3,b,b_3)S_t(x_3)\big\}\;,
\\
M^{LL}_{A\sigma} &=& 32\pi C_F m^4_B/\sqrt{2N_c} \int dx_B dz dx_3\int b_B db_B b_3 db_3\phi_B(x_B,b_B)
\big\{\big[(\eta-1)[x_3(1-\eta)+x_B +\eta(1-z)]\phi^A\phi^0
\nonumber\\
&+& r_0\sqrt\eta(x_3(1-\eta)+x_B+\eta)(\phi^P+\phi^T)(\phi^s-\phi^t)+ r_0\sqrt\eta(1-z)
(\phi^P-\phi^T)(\phi^s+\phi^t)
\nonumber\\
&+&2r_0\sqrt\eta(\phi^P\phi^s+\phi^T\phi^t)\big]
 E_{1gh}(t_{1g})h_{1g}(x_B,z,x_3,b_B,b_3)+\big[(1-\eta^2)(1-z)\phi^A\phi^0
\nonumber\\
&+&r_0\sqrt\eta(x_B-x_3(1-\eta)-\eta)(\phi^P-\phi^T)(\phi^s+\phi^t)
-r_0\sqrt\eta(1-z)(\phi^P+\phi^T)(\phi^s-\phi^t)\big]
\nonumber\\
&\times& E_{1gh}(t_{1h})h_{1h}(x_B,z,x_3,b_B,b_3)\big\}\;,
\\
M^{LR}_{A\sigma} &=& -32\pi C_F m^4_B/\sqrt{2N_c} \int dx_B dz dx_3\int b_B db_B b_3 db_3
\phi_B(x_B,b_B)\big\{\big[\sqrt\eta(1-\eta)(1+z)\phi^A(\phi^s-\phi^t)
\nonumber\\
&+&r_0(2-x_B-x_3(1-\eta))(\phi^P+\phi^T)\phi^0 +r_0\eta(z\phi^P-(2+z)\phi^T)\phi^0\big]
E_{1gh}(t_{1g})h_{1g}(x_B,z,x_3,b_B,b_3)
\nonumber\\
&+& \big[\sqrt\eta(1-\eta)(1-z)\phi^A
(\phi^s-\phi^t)+r_0((x_3(1-\eta)-x_B)(\phi^P+\phi^T)
+\eta((2-z)\phi^P+z\phi^T))\phi^0 \big]
\nonumber\\
&\times&E_{1gh}(t_{1h})h_{1h}(x_B,z,x_3,b_B,b_3)\big\}\;,
\\
M^{SP}_{A\sigma} &=&32\pi C_F m^4_B/\sqrt{2N_c} \int dx_B dz dx_3\int b_B db_B b_3 db_3
\phi_B(x_B,b_B)\big\{\big[\phi^0\phi^A(\eta-1)(z(1+\eta)-1)
\nonumber\\
&+&r_0\sqrt{\eta}(\phi^T-\phi^P)(\phi^s+\phi^t)((\eta-1)(1-x_3)+x_B)+r_0\sqrt{\eta}(\phi^T+\phi^P)(\phi^s-\phi^t)z -4r_0\sqrt{\eta}\phi^P\phi^s \big]
\nonumber\\
&\times& E_{1gh}(t_{1g})h_{1g}(x_B,z,x_3,b_B,b_3) + \big[\phi^0\phi^A(\eta-1)(x3(1-\eta)+\eta(2-z)-x_B)
\nonumber\\
&+&(\phi^T+\phi^P)(\phi^s-\phi^t)r_0\sqrt{\eta}(x_3(1-\eta)+\eta-x_B)+(\phi^P-\phi^T)(\phi^s+\phi^t)r_0\sqrt{\eta} (1-z) \big]
\nonumber\\
&\times&E_{1gh}(t_{1h})h_{1h}(x_B,z,x_3,b_B,b_3)\big\}\;,
\\
F^{SP}_{TP}&=&16\pi C_F F_s(\omega^2)\sqrt{\eta} m^4_B \int dx_B dx_3\int b_B db_B b_3 db_3 \phi_B(x_B,b_B)
\big\{\big[(\eta-1)\phi^A + r_0x_3(\eta-1)(\phi^P-\phi^T)
\nonumber\\
&-&2r_0\phi^P\big]E_{2ab}(t_{2a})h_{2a}(x_B,x_3,b_B,b_3)S_t(x_3)
+\left[x_B(\eta-1)\phi^A+2r_0(\eta+x_B-1)\phi^P\right]
\nonumber\\
&\times& E_{2ab}(t_{2b})h_{2b}(x_B,x_3,b_B,b_3)S_t(x_B) \big\}\;,
\\
M^{LL}_{TP}&=&32\pi C_F m^4_B/\sqrt{2N_c} \int dx_B dz dx_3\int b_B db_B b db
\phi_B(x_B,b_B)\phi^0 \big\{\big[(1-x_B-z)(1-\eta^2)\phi^A
\nonumber\\
&-&r_0x_3(1-\eta)(\phi^P-\phi^T)+r_0(x_B+z)\eta(\phi^P+\phi^T)
-2r_0\eta\phi^P\big] E_{2cd}(t_{2c})h_{2c}(x_B,z,x_3,b_B,b)
\nonumber\\
&-&\big[(z-x_B+x_3(1-\eta))(1-\eta)\phi^A+r_0(x_B-z)\eta(\phi^P-\phi^T)
-r_0x_3(1-\eta)(\phi^P+\phi^T) \big]
\nonumber\\
&\times&E_{2cd}(t_{2d})h_{2d}(x_B,z,x_3,b_B,b)\big\}\;,
\\
M^{LR}_{TP}&=&32\pi C_F m^4_B\sqrt{\eta}/\sqrt{2N_c} \int dx_B dz dx_3\int b_B db_B b db
\phi_B(x_B,b_B) \big\{\big[(1-x_B-z)(1-\eta)(\phi^s+\phi^t)\phi^A
\nonumber\\
&+& r_0(1-x_B-z)(\phi^s+\phi^t)(\phi^P-\phi^T)+r_0(x_3(1-\eta)+\eta)(\phi^s-\phi^t)
(\phi^P+\phi^T)\big]
\nonumber\\
&\times& E_{2cd}(t_{2c})h_{2c}(x_B,z,x_3,b_B,b)
-\big[(z-x_B)(1-\eta)(\phi^s-\phi^t)\phi^A+r_0(z-x_B)(\phi^s-\phi^t)(\phi^P-\phi^T)
\nonumber\\
&+&r_0x_3(1-\eta)(\phi^s+\phi^t)(\phi^P+\phi^T)\big]
E_{2cd}(t_{2d}) h_{2d}(x_B,z,x_3,b_B,b)\big\}\;,
\\
M^{SP}_{TP} &=&32\pi C_F m^4_B/\sqrt{2N_c} \int dx_B dz dx_3\int b_B db_B b db \phi_B(x_B,b_B)\phi^0
\big\{\big[(1+\eta-x_B-z+x_3(1-\eta))(1-\eta)\phi^A
\nonumber\\
&+&r_0\eta(x_B+z)(\phi^P-\phi^T) -r_0x_3(1-\eta)(\phi^P+\phi^T)
-2r_0\eta\phi^P\big] E_{2cd}(t_{2c})h_{2c}(x_B,z,x_3,b_B,b)
\nonumber\\
&-&\left[(z-x_B)(1-\eta^2)\phi^A-r_0x_3(1-\eta)(\phi^P-\phi^T)
+r_0\eta(x_B-z)(\phi^P+\phi^T)\right]
\nonumber\\
&\times& E_{2cd}(t_{2d})h_{2d}(x_B,z,x_3,b_B,b)\big\}\;,
\\
F^{LL}_{AP} &=& 8\pi C_F m^4_B f_B \int dz dx_3\int b db b_3 db_3
\big\{\big[(x_3(1-\eta)-1)(1-\eta)\phi^A\phi^0 +2r_0\sqrt\eta(x_3(1-\eta)(\phi^P-\phi^T)
\nonumber\\
&-&2\phi^P)\phi^s\big] E_{2ef}(t_{2e})h_{2e}(z,x_3,b,b_3)S_t(x_3)
+\big[z(1-\eta)\phi^A\phi^0+2r_0\sqrt\eta\phi^P((1-\eta)(\phi^s-\phi^t)
\nonumber\\
&+&z(\phi^s+\phi^t))\big] E_{2ef}(t_{2f})h_{2f}(z,x_3,b,b_3)S_t(z) \big\}\;,
\\
F^{SP}_{AP} &=& 16\pi C_F m^4_B f_B  \int dz dx_3\int b db b_3 db_3
\big\{\big[2\sqrt\eta(1-\eta)\phi^A\phi^s +r_0(1-x_3)(\phi^P+\phi^T)\phi^0
\nonumber\\
&+&r_0\eta((1+x_3)\phi^P-(1-x_3)\phi^T)\phi^0\big] E_{2ef}(t_{2e})h_{2e}(z,x_3,b,b_3)S_t(x_3)
\nonumber\\
&+&\left[2r_0(1-\eta(1-z))\phi^P\phi^0+z\sqrt\eta((1-\eta)\phi^A(\phi^s-\phi^t) \right]
E_{2ef}(t_{2f})h_{2f}(z,x_3,b,b_3)S_t(z)\big\}\;,
\\
M^{LL}_{AP}&=&32\pi C_F m^4_B/\sqrt{2N_c} \int dx_B dz dx_3\int b_B db_B b_3 db_3
\phi_B(x_B,b_B) \big\{\big[(\eta-1)(-\eta+(1+\eta)(x_B+z))\phi^A\phi^0
\nonumber\\
&+&r_0\sqrt\eta(x_3(1-\eta)+\eta)(\phi^P+\phi^T)(\phi^s-\phi^t)
+r_0\sqrt\eta(1-x_B-z)(\phi^P-\phi^T)(\phi^s+\phi^t)-4r_0\sqrt\eta\phi^P\phi^s\big]
\nonumber\\
&\times&E_{2gh}(t_{2g})h_{2g}(x_B,z,x_3,b_B,b_3)+\big[(1-\eta)((1-x_3)(1-\eta)-\eta(x_B-z))\phi^A\phi^0
-r_0\sqrt\eta(x_B-z)
\nonumber\\
&\times&(\phi^P+\phi^T)(\phi^s-\phi^t)+r_0\sqrt\eta(1-\eta)(1-x_3)(\phi^P-\phi^T)(\phi^s+\phi^t)\big]
E_{2gh}(t_{2h})h_{2h}(x_B,z,x_3,b_B,b_3) \big\}\;,
\\
M^{LR}_{AP}&=&32\pi C_F m^4_B/\sqrt{2N_c} \int dx_B dz dx_3\int b_B db_B b_3 db_3
\phi_B(x_B,b_B)\big\{\big[\sqrt\eta(1-\eta)(2-x_B-z)\phi^A(\phi^s+\phi^t)
\nonumber\\
&-&r_0(1+x_3)(\phi^P-\phi^T)\phi^0 -r_0\eta[(1-x_B-z)(\phi^P+\phi^T)
-x_3(\phi^P-\phi^T)+2\phi^P]\phi^0\big]
\nonumber\\
&\times&E_{2gh}(t_{2g})h_{2g}(x_B,z,x_3,b_B,b_3)-\big[r_0(1-\eta)(x_3- 1)(\phi^P-\phi^T)\phi^0+\sqrt\eta(x_B-z)[r_0\sqrt\eta(\phi^P+\phi^T)\phi^0
\nonumber\\
&+&(1-\eta)\phi^A(\phi^s+\phi^t)] \big]E_{2gh}(t_{2h}) h_{2h}(x_B,z,x_3,b_B,b_3)\big\}\;,
\\
M^{SP}_{AP}&=&32\pi C_F m^4_B/\sqrt{2N_c} \int dx_B dz dx_3\int b_B db_B b_3 db_3
\phi_B(x_B,b_B)\big\{\big[(\eta-1)\phi^0\phi^A(\eta(x_3+x_B+z-2)
\nonumber\\
&-&x_3+1)+r_0\sqrt{\eta}[(1-\eta)(1-x_3)(\phi^P-\phi^T)(\phi^s+\phi^t) +(z+x_B)(\phi^P+\phi^T)(\phi^s-\phi^t)
\nonumber\\
&+&2(\phi^P\phi^s+\phi^T\phi^t)]\big]
E_{2gh}(t_{2g})h_{2g}(x_B,z,x_3,b_B,b_3)-\big[r_0\sqrt{\eta}(\phi^P+\phi^T)(\phi^s-\phi^t)(1-\eta)(x_3-1)
\nonumber\\
&+&r_0\sqrt{\eta}(\phi^P-\phi^T)(\phi^s+\phi^t)(x_B-z)+\phi^0\phi^A(1-\eta^2)(x_B-z) \big]E_{2gh}(t_{2h}) h_{2h}(x_B,z,x_3,b_B,b_3)\big\}\;.
\end{eqnarray}
where the hard functions are defined as
\begin{eqnarray}
h_i(x_1,x_2,(x_3,)b_1,b_2)&=&h_1(\beta,b_2)\times h_2(\alpha,b1,b_2)  \nonumber\\
h_1(\beta,b_2)&=&\left\{ \begin{array}{ll} K_0(\sqrt{\beta} b_2), \quad &\beta \geq 0, \\
              \frac{i\pi}{2}H^{(1)}_0(\sqrt{-\beta} b_2), \quad &\beta < 0,\end{array} \right.  \nonumber\\
h_2(\alpha,b1,b_2)&=& \left\{ \begin{array}{ll} \theta(b_2-b_1) K_0(\sqrt{\alpha}b_2)I_0(\sqrt{\alpha}b_1), \quad &\alpha \geq 0, \\
              \theta(b_2-b_1) \frac{i\pi}{2}H^{(1)}_0(\sqrt{-\alpha}b_2)J_0(\sqrt{-\alpha}b_1), \quad &\alpha < 0,\end{array} \right.
\label{3-bodyhardfunction}
\end{eqnarray}
where $E_{1mn},E_{2mn},E_{3mn},E_{4mn}$($m=a,c,e,g$ and $n=b,d,f,h$) are the evolution factors,
\begin{eqnarray}
E_{1ab}(t)&=&\alpha(t)\exp[-S_B(t)-S_{\sigma}(t)],   \nonumber\\
E_{1cd}(t)&=&\alpha(t)\exp[-S_B(t)-S_{\sigma}(t)-S_P(t)]_{b=b_B},  \nonumber\\
E_{1ef}(t)&=&\alpha(t)\exp[-S_P(t)-S_{\sigma}(t)],   \nonumber\\
E_{1gh}(t)&=&\alpha(t)\exp[-S_B(t)-S_{\sigma}(t)-S_P(t)]_{b=b_3}, \nonumber\\
E_{2ab}(t)&=&\alpha(t)\exp[-S_B(t)-S_P(t)],   \nonumber\\
E_{2cd}(t)&=&\alpha(t)\exp[-S_B(t)-S_{\sigma}(t)-S_P(t)]_{b_3=b_B},  \nonumber\\
E_{2ef}(t)&=&E_{1ef}(t),  \nonumber\\
E_{2gh}(t)&=&E_{1gh}(t),  \nonumber\\
E_{3ab}(t)&=&\alpha(t)\exp[-S_B(t)-S_{\sigma}(t)],   \nonumber\\
E_{3cd}(t)&=&\alpha(t)\exp[-S_B(t)-S_{\sigma}(t)-S_D(t)]_{b=b_B},  \nonumber\\
E_{3ef}(t)&=&\alpha(t)\exp[-S_D(t)-S_{\sigma}(t)],   \nonumber\\
E_{3gh}(t)&=&\alpha(t)\exp[-S_B(t)-S_{\sigma}(t)-S_D(t)]_{b=b_3}, \nonumber\\
E_{4ef}(t)&=&E_{3ef}(t),  \nonumber\\
E_{4gh}(t)&=&E_{3gh}(t)\,.
\end{eqnarray}

The Sudakov form factors are defined as
\begin{eqnarray}
&& S_B(t)=s\left(\frac{x_B m_B}{\sqrt2},b_B\right)+\frac{5}{3}\int^{t}_{1/b_B}\frac{d\bar{\mu}}{\bar{\mu}}\gamma_q(\alpha_s(\bar{\mu})), \nonumber\\
&& S_{\sigma}(t)=s\left(\frac{z(1-r^2) m_B}{\sqrt2},b\right)+s\left(\frac{(1-z)(1-r^2) m_B}{\sqrt2},b\right) + 2\int^{t}_{1/b}\frac{d\bar{\mu}}{\bar{\mu}}\gamma_q(\alpha_s(\bar{\mu})),\nonumber\\
&& S_{D,P}(t)=s\left(\frac{x_3 m_B}{\sqrt2},b_3\right)+s\left(\frac{(1-x_3) m_B}{\sqrt2},b_3\right)+ 2\int^{t}_{1/b_3} \frac{d\bar{\mu}}{\bar{\mu}} \gamma_q(\alpha_s(\bar{\mu}))\,.
\end{eqnarray}

Practically, the mentioned hard scales are chosen as
\begin{align}
&t_{1a}= {\rm Max} \left\{\sqrt{\vert \alpha_{1a} \vert},\sqrt{\vert\beta_{1a}\vert},1/b_B,1/b\right\},
&&t_{1b}= {\rm Max} \left\{\sqrt{\vert\alpha_{1b}\vert},\sqrt{\vert \beta_{1b} \vert},1/b_B,1/b\right\},\nonumber\\
&t_{1c}= {\rm Max} \left\{\sqrt{\vert\alpha_{1c}\vert},\sqrt{\vert \beta_{1c} \vert},1/b_B,1/b_3\right\},
&&t_{1d}= {\rm Max} \left\{\sqrt{\vert\alpha_{1d}\vert},\sqrt{\vert \beta_{1d} \vert},1/b_B,1/b_3\right\},\nonumber\\
&t_{1e}= {\rm Max} \left\{\sqrt{\vert\alpha_{1e}\vert},\sqrt{\vert \beta_{1e} \vert},1/b_3,1/b\right\},
&&t_{1f}= {\rm Max} \left\{\sqrt{\vert\alpha_{1f}\vert},\sqrt{\vert \beta_{1f} \vert},1/b_3,1/b\right\},\nonumber\\
&t_{1g}= {\rm Max} \left\{\sqrt{\vert\alpha_{1g}\vert},\sqrt{\vert \beta_{1g} \vert},1/b_B,1/b_3\right\},
&&t_{1h}= {\rm Max} \left\{\sqrt{\vert\alpha_{1h}\vert},\sqrt{\vert \beta_{1h} \vert},1/b_B,1/b_3\right\}, \nonumber\\
&t_{2a}= {\rm Max} \left\{\sqrt{\vert\alpha_{2a}\vert},\sqrt{\vert \beta_{2a} \vert},1/b_B,1/b_3\right\},
&&t_{2b}= {\rm Max} \left\{\sqrt{\vert\alpha_{2b}\vert},\sqrt{\vert \beta_{2b} \vert},1/b_B,1/b_3\right\},\nonumber\\
&t_{2c}= {\rm Max} \left\{\sqrt{\vert\alpha_{2c}\vert},\sqrt{\vert \beta_{2c} \vert},1/b_B,1/b\right\},
&&t_{2d}= {\rm Max} \left\{\sqrt{\vert\alpha_{2d}\vert},\sqrt{\vert \beta_{2d} \vert},1/b_B,1/b\right\},\nonumber\\
&t_{2e}= {\rm Max} \left\{\sqrt{\vert\alpha_{2e}\vert},\sqrt{\vert \beta_{2e} \vert},1/b_3,1/b\right\},
&&t_{2f}= {\rm Max} \left\{\sqrt{\vert\alpha_{2f}\vert},\sqrt{\vert \beta_{2f} \vert},1/b_3,1/b\right\},\nonumber\\
&t_{2g}= {\rm Max} \left\{\sqrt{\vert\alpha_{2g}\vert},\sqrt{\vert \beta_{2g} \vert},1/b_B,1/b_3\right\},
&&t_{2h}= {\rm Max} \left\{\sqrt{\vert\alpha_{2h}\vert},\sqrt{\vert \beta_{2h} \vert},1/b_B,1/b_3\right\},\nonumber\\
&t_{3a}= {\rm Max} \left\{\sqrt{\vert \alpha_{3a} \vert},\sqrt{\vert\beta_{3a}\vert},1/b_B,1/b\right\},
&&t_{3b}= {\rm Max} \left\{\sqrt{\vert\alpha_{3b}\vert},\sqrt{\vert \beta_{3b} \vert},1/b_B,1/b\right\},\nonumber\\
&t_{3c}= {\rm Max} \left\{\sqrt{\vert\alpha_{3c}\vert},\sqrt{\vert \beta_{3c} \vert},1/b_B,1/b_3\right\},
&&t_{3c}^{\prime}= {\rm Max} \left\{\sqrt{\vert\alpha_{3c}\vert},\sqrt{\vert \beta^{\prime}_{1c} \vert},1/b_B,1/b_3\right\},\nonumber\\
&t_{3d}= {\rm Max} \left\{\sqrt{\vert\alpha_{3d}\vert},\sqrt{\vert \beta_{3d} \vert},1/b_B,1/b_3\right\},
&&t_{3d}^{\prime}= {\rm Max} \left\{\sqrt{\vert\alpha_{3d}\vert},\sqrt{\vert \beta^{\prime}_{1d} \vert},1/b_B,1/b_3\right\},\nonumber\\
&t_{4e}= {\rm Max} \left\{\sqrt{\vert\alpha_{4e}\vert},\sqrt{\vert \beta_{4e} \vert},1/b_3,1/b\right\},
&&t_{4f}= {\rm Max} \left\{\sqrt{\vert\alpha_{4f}\vert},\sqrt{\vert \beta_{4f} \vert},1/b_3,1/b\right\},\nonumber\\
&t_{4g}= {\rm Max} \left\{\sqrt{\vert\alpha_{4g}\vert},\sqrt{\vert \beta_{4g} \vert},1/b_B,1/b_3\right\},
&&t_{4h}= {\rm Max} \left\{\sqrt{\vert\alpha_{4h}\vert},\sqrt{\vert \beta_{4h} \vert},1/b_B,1/b_3\right\}. \label{Eq:A23}
\end{align}
The jet function $S_t$ resums the threshold double logarithm and can be parameterized as
\begin{equation}
\frac{2^{1+2c}\Gamma(3/2+c)}{\sqrt{\pi}\Gamma(1+c)}[x(1-x)]^c\;,
\end{equation}
with $c=0.4$ for numerical calculation. The parameters in Eq. \eqref{Eq:A23} take the form of:
\begin{eqnarray}
\alpha_{1a}&=&z m_B^2 \nonumber\\
\beta_{1a}&=&x_B z m_B^2=\beta_{1b}=\alpha_{1c}=\alpha_{1d} \nonumber\\
\alpha_{1b}&=&(x_B-\eta)m_B^2      \nonumber\\
\beta_{1c}&=&-z[(1-\eta)(1-x_3)-x_B]m_B^2 \nonumber\\
\beta_{1d}&=&-z[(1-\eta)x_3-x_B]m_B^2 \nonumber\\
\alpha_{1e}&=&(z-1)m_B^2  \nonumber\\
\beta_{1e}&=&-(1-z)[\eta+(1-\eta)x_3]m_B^2=\beta_{1f}=\alpha_{1g}=\alpha_{1h} \nonumber\\
\alpha_{1f}&=&-(\eta+(1-\eta)x_3)m_B^2                 \nonumber\\
\beta_{1g}&=&\{1-z[(1-\eta)(1-x_3)-x_B]\}m_B^2 \nonumber\\
\beta_{1h}&=&-(1-z)[(1-\eta)x_3+\eta-x_B]m_B^2\nonumber\\
\alpha_{2a}&=&x_3(1-\eta) m_B^2 \nonumber\\
\beta_{2a}&=&x_3x_B(1-\eta) m_B^2=\beta_{2b}=\alpha_{2c}=\alpha_{2d} \nonumber\\
\alpha_{2b}&=&x_B(1-\eta)m_B^2      \nonumber\\
\beta_{2c}&=&(1-x_B-z)[(\eta-1)x_3-\eta]m_B^2 \nonumber\\
\beta_{2d}&=&(1-\eta)(x_B-z)m_B^2 \nonumber\\
\alpha_{2e}&=&-[1-x_3(1-\eta)]m_B^2                 \nonumber\\
\beta_{2e}&=&-z(1-\eta)(1-x_3)m_B^2=\beta_{2f}=\alpha_{2g}=\alpha_{2h} \nonumber\\
\alpha_{2f}&=&z(\eta-1)m_B^2                 \nonumber\\
\beta_{2g}&=&\{1-(1-z-x_B)[\eta+(1-\eta)x_3]\}m_B^2 \nonumber\\
\beta_{2h}&=&(x_B-z)(1-\eta)(1-x_3)m_B^2  \nonumber\\
\alpha_{3a}&=&z(1-r^2)m_B^2 \nonumber\\
\beta_{3a}&=&x_B z (1-r^2)m_B^2=\beta_{3b}=\alpha_{3c}=\alpha_{3d}=\alpha^{\prime}_{3c}=\alpha^{\prime}_{3d} \nonumber\\
\alpha_{3b}&=&(1-r^2)(x_B-\eta)m_B^2      \nonumber\\
\beta_{3c}&=&-[z(1-r^2)+r^2(1-x_3)][(1-\eta)(1-x_3)-x_B]m_B^2 \nonumber\\
\beta_{3d}&=&\{r_c^2-[z(1-r^2)+x_3 r^2][(1-\eta)x_3-x_B]\}m_B^2 \nonumber\\
\beta^{\prime}_{3c}&=&\{r_c^2-[z(1-r^2)+r^2(1-x_3)][(1-\eta)(1-x_3)-x_B]\}m_B^2 \nonumber\\
\beta^{\prime}_{3d}&=&-[z(1-r^2)+x_3 r^2][(1-\eta)x_3-x_B]m_B^2 \nonumber\\
\alpha_{3e}&=&-[1-z(1-r^2)-r_c^2]m_B^2                   \nonumber\\
\beta_{3e}&=&-[(1-r^2)(1-z)+x_3r^2][\eta+(1-\eta)x_3]m_B^2=\beta_{3f}=\alpha_{3g}=\alpha_{3h} \nonumber\\
\alpha_{3f}&=&-(1-r^2+x_3r^2)(\eta+(1-\eta)x_3)m_B^2                 \nonumber\\
\beta_{3g}&=&\{1-[z(1-r^2)+r^2(1-x_3)][(1-\eta)(1-x_3)-x_B]\}m_B^2 \nonumber\\
\beta_{3h}&=&-[(1-z)(1-r^2)+r^2x_3][(1-\eta)x_3+\eta-x_B]m_B^2\nonumber\\
\alpha_{4e}&=&-(1-x_3r^2)[1-x_3(1-\eta)]m_B^2                 \nonumber\\
\beta_{4e}&=&-[(1-x_3)r^2+z(1-r^2)](1-\eta)(1-x_3)m_B^2=\beta_{4f}=\alpha_{4g}=\alpha_{4h} \nonumber\\
\alpha_{4f}&=&\{r_c^2-[r^2+z(1-r^2)](1-\eta)\}m_B^2                 \nonumber\\
\beta_{4g}&=&\{1-[(1-r^2)(1-z)+x_3r^2-x_B][\eta+(1-\eta)x_3]\}m_B^2 \nonumber\\
\beta_{4h}&=&-[r^2(1-x_3)+z(1-r^2)-x_B](1-\eta)(1-x_3)m_B^2
\end{eqnarray}

\bibliography{BPf}
\bibliographystyle{apsrev4-1}

\end{document}